\documentstyle[aps,preprint]{revtex}
\begin{document}
\draft
\title{Anomalous superconducting state gap size vs 
$T_c$ behavior in underdoped Bi$_2$Sr$_2$Ca$_{1-
x}$Dy$_x$Cu$_2$O$_{8+\delta}$}
\author{J. M. Harris, Z.-X. Shen, P. J. White, 
D. S. Marshall and M. C. Schabel}
\address{Department of Applied Physics and Stanford Synchrotron 
Radiation Laboratory, Stanford University, 
Stanford, CA 94305-4045}
\author{J. N. Eckstein and I. Bozovic}
\address{Ginzton Research Center, Varian Associates, Palo Alto, 
California 94304-1025}
\date{July 30, 1996}
\maketitle

\begin{abstract}
We report angle-resolved photoemission spectroscopy 
measurements of the 
excitation gap in underdoped superconducting thin films of 
Bi$_2$Sr$_2$Ca$_{1-x}$Dy$_x$Cu$_2$O$_{8+\delta}$.  
As $T_c$ is reduced by a factor of 2 by underdoping, the 
superconducting state 
gap $\Delta$ does not fall proportionally, but instead stays 
constant or increases slightly, 
in violation of the 
BCS mean-field theory result.  The different doping dependences 
of $\Delta$ and $kT_c$ 
indicate that they represent different energy scales.  The 
measurements also 
show that $\Delta$ is highly anisotropic and consistent with a 
$d_{x^2-y^2}$ order parameter, 
as in previous studies of samples with higher dopings.  However, 
in these 
underdoped samples, the anisotropic gap persists well above 
$T_c$\@.  The existence of a normal state gap is related to the 
failure of $\Delta$ to scale with $T_c$ in theoretical 
models that predict pairing without
phase coherence above $T_c$\@.  
\end{abstract}

The superconducting state of a metal is characterized 
by an energy gap in the spectrum of electronic 
excitations.  BCS-Eliashberg mean-field theory \cite{Schrieffer} 
has been used 
successfully for decades in describing the superconducting 
state gap $\Delta$ in 
conventional superconductors such as Al or Nb.
The typical mean-field theory result that the superconducting 
transition temperature $T_c$ is proportional to 
$\Delta$ has been abundantly confirmed for 
conventional superconductors.  Here we report data from 
angle-resolved photoemission spectroscopy showing that 
this proportionality is violated in underdoped samples of 
the high temperature superconductor (HTSC) 
Bi$_2$Sr$_2$Ca$_{1-x}$Dy$_x$Cu$_2$O$_{8+\delta}$,
implying a novel transition into the superconducting state.
	
A large variety of 
experimental measurements in underdoped HTSC have shown 
evidence for a suppression 
in the intensity of low energy excitations {\it above} $T_c$, 
including NMR \cite{NMR}, 
resistivity \cite{Ito,Bucher}, 
specific heat \cite{Loram}, and {\it c}-axis optical 
conductivity \cite{Homes}.  
The relation, if any, between the superconducting state 
gap and this pseudogap 
or normal state gap is still an open question.  
Angle-resolved photoemission spectroscopy (ARPES) is uniquely 
able to measure the gap 
magnitude anisotropy.  Measurements on slightly overdoped 
Bi$_2$Sr$_2$CaCu$_2$O$_{8+\delta}$ (BSCCO) 
below $T_c$ are consistent with the four-lobed $d_{x^2-y^2}$ gap 
\cite{Shen} 
and show no normal state gap (or a very small one).  
The present results on {\it underdoped} BSCCO samples with a 
wide range of transition temperatures show the 
same strong angular dependence of the superconducting state gap, 
but in addition they 
show a {\it normal state }Êgap with very similar magnitude and 
angular dependence, in 
agreement with other recent ARPES work \cite{Marshall,Loeser,DingTdep}.  
The normal state gap persists to temperatures well above $T_c$\@.  
These observations can be related to our finding that the 
superconducting state gap fails to 
scale with $T_c$ as expected from mean-field theory.  We discuss 
below how models with 
phase fluctuations in the order parameter predict pairing in the 
normal state of underdoped 
HTSC up to a characteristic mean-field temperature $T_{MF}$, 
giving the two gaps a common 
origin and accounting for the non-mean-field gap size vs $T_c$ 
dependence.  In these models, 
$T_c$ is the temperature for phase coherence between pairs rather 
than the onset of pairing.

The angle-resolved gap measurements were carried out using a 
Scienta hemispherical
analyzer operating with 20 meV FWHM resolution ($\sigma=9$ 
meV) as measured by a Au 
reference Fermi edge.  The photon energy was 21.2 eV, and the 
base pressure of the 
vacuum system was $5\times 10^{-11}$ torr.  The analyzer 
acceptance angle 
was $\pm 1^\circ$, corresponding to a 
$k$-space window of radius 0.045$\pi/a$ or 0.037 \AA$^{-1}$.  
Our BSCCO samples were untwinned 
thin film crystals \cite{EcksteinAPL} grown by atomic 
layer-by-layer molecular-beam 
epitaxy \cite{EcksteinARMS} and an underdoped bulk single crystal.  
Underdoping was accomplished by substituting trivalent 
Dy for divalent Ca and/or reducing the oxygen content.  The nine 
thin-film samples 
measured were cut from three larger samples with $T_c=46$, 78, 
and 89 K, respectively, determined 
from $\rho$ vs $T$ superconducting transition midpoints.  The 
samples were cleaved in 
ultrahigh vacuum using a top-post method.  The thin films give 
results that are similar to 
those from bulk single crystals for the same hole 
doping \cite{filmnote}.

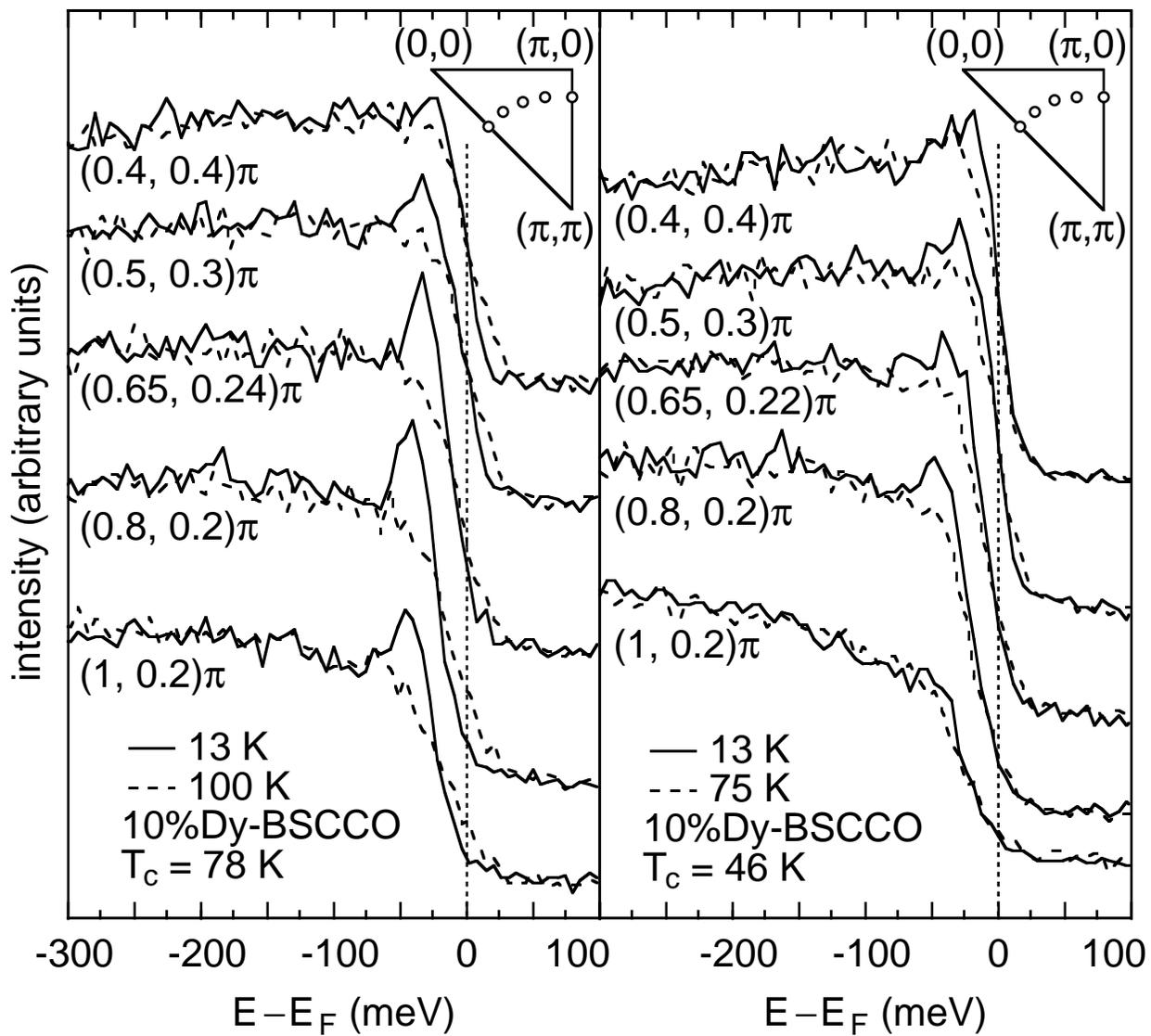
\begin{figure}
\caption{ARPES spectra near the Fermi energy for underdoped 
single crystal thin films of 
Bi$_2$Sr$_2$Ca$_{1-x}$Dy$_x$Cu$_2$O$_{8+\delta}$ in the 
superconducting and normal states.  
$k$-space positions 
were selected on the underlying FS to facilitate measuring 
the energy gap.  
The systematic shift of leading edge position with $k$ shows an 
anisotropic energy gap.
}
\label{fig1}
\end{figure}

Raw data on underdoped BSCCO of two different oxygen dopings are 
plotted in Fig.\ \ref{fig1} for
temperatures well below and well above $T_c$\@.  The data 
are taken as energy distribution curves 
(EDC's) consisting of photoemission intensity vs energy at fixed 
{\bf k}.  The superconducting 
state gap magnitude $|\Delta_k|$ can be found using the usual 
expression for excitations from the 
superconducting ground state, $E_k=\surd (\varepsilon_k^2 + 
|\Delta_k|^2)$, 
where $\varepsilon_k$ is the normal state (or ungapped) 
single-particle energy relative to $E_F$.  At the Fermi 
surface (FS), $\varepsilon_k = 0$ by definition, 
and the gap is simply the minimum energy excitation.  In ARPES 
the gap can be found by 
stepping through {\it k}-space along a particular direction and 
finding where the quasiparticle 
feature is closest to $E_F$.  

Data from BSCCO are more difficult to interpret than data from 
conventional materials 
since the EDC line shapes are broad and are not currently well 
understood.  The samples 
that are more strongly underdoped have even broader features, 
especially in the vicinity
of ($\pi$,0) in the normal state.  In order to avoid the 
ambiguities of picking a particular 
spectral function and background to fit the line shapes, we instead 
characterize the gap using 
the midpoint energy of the leading edge 
(the edge nearest $E_F$ or 0 in Fig.\ \ref{fig1}) as in previous 
work \cite{Shen}.  As in optimally-doped BSCCO \cite{Ding}, gap 
values below $T_c$ 
from leading edge 
shifts were found to be somewhat smaller than those extracted from 
fits to a broadened 
BCS spectral function; however, we stress that the trends in 
our data are insensitive to the 
specific method of characterizing the gap.
	
The five {\it k}-space positions shown (Fig.\ \ref{fig1} inset) 
were chosen by taking five cuts in 
the Brillouin zone and selecting {\bf k} so that the leading 
edge is closest in energy to $E_F$ (i.e. on 
the underlying FS).  It is evident that the leading edge 
positions change 
monotonically with angle, and that the superconducting and normal 
state leading edge positions 
are quite similar.  Characteristic narrow peaks appear in the 
superconducting state near $E_F$.  
They are especially clear away from the 0.4($\pi$,$\pi$) Fermi 
surface crossing.  The observation 
of the narrow peaks indicates that the broad normal state 
line shape is intrinsic to the doping 
level and not primarily the effect of disorder or impurity 
scattering, since the scattering 
would also be expected to broaden the features below $T_c$.
Also, we note a strong and systematic increase in the 
low-temperature height of the peaks with doping in the 
underdoped regime.

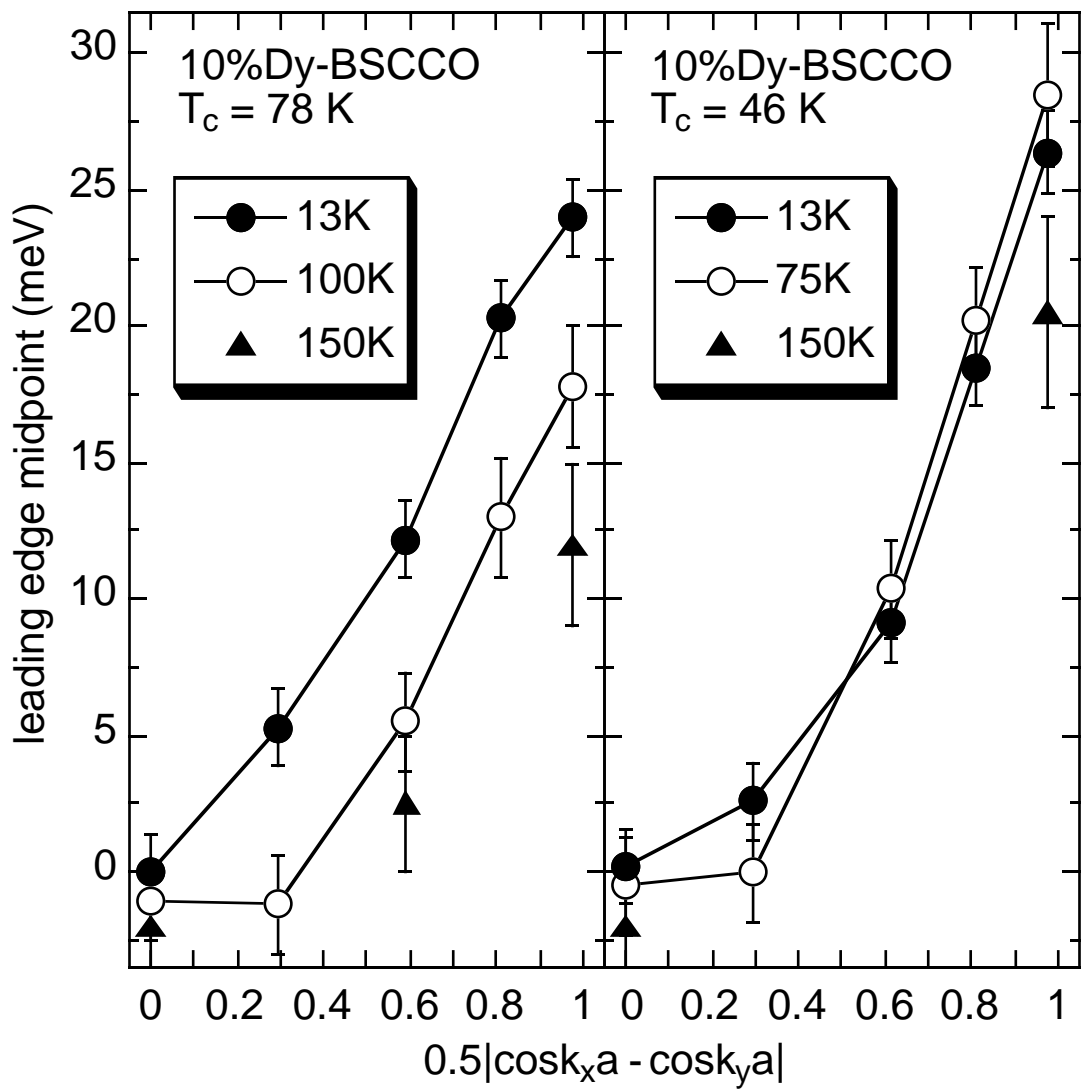
\begin{figure}
\caption{Leading edge midpoint shifts from $E_F$, indicative of an 
anisotropic energy gap, in 
the superconducting and normal states of Bi$_2$Sr$_2$Ca$_{1-
x}$Dy$_x$Cu$_2$O$_{8+\delta}$ extracted from the 
ARPES spectra of Fig. 1.  The abscissa, $0.5|\cos k_xa - \cos 
k_ya|$, was selected for comparison 
to a $d_{x^2-y^2}$ gap, which would be a straight line on this 
plot.  The ``dirty $d$-wave'' scenario 
predicts flattening near the origin.}
\label{fig2}
\end{figure}

The leading edge midpoints of the EDCs of Fig.\ \ref{fig1} are 
plotted vs $0.5|\cos k_xa - 
\cos k_ya|$ in Fig.\ \ref{fig2}.  On this plot the $d_{x^2-y^2}$ 
gap prediction 
is a straight line that passes 
through the origin.  The error bars are a combination of 
uncertainty in the leading edge 
midpoint position and uncertainty in $E_F$ ($\pm$1 meV).  The 
data agree reasonably well with the 
strong anisotropy of the $d$-wave gap, but in some cases the 
measured values show a flattening 
near the {\it d}-wave node position (i.e., the origin in Fig.\ 
\ref{fig2}).  One explanation is the ``dirty 
$d$-wave'' scenario \cite{Borkowski,Fehrenbacher} where 
impurity scattering broadens the point node into a region 
of finite width.  Systematic impurity doping studies using ARPES 
would be useful to clarify 
this point.  Another contribution to the flattening is finite 
{\it k} resolution combined with the 
strong energy dispersion in the node direction 
\cite{Fehrenbachersub}.

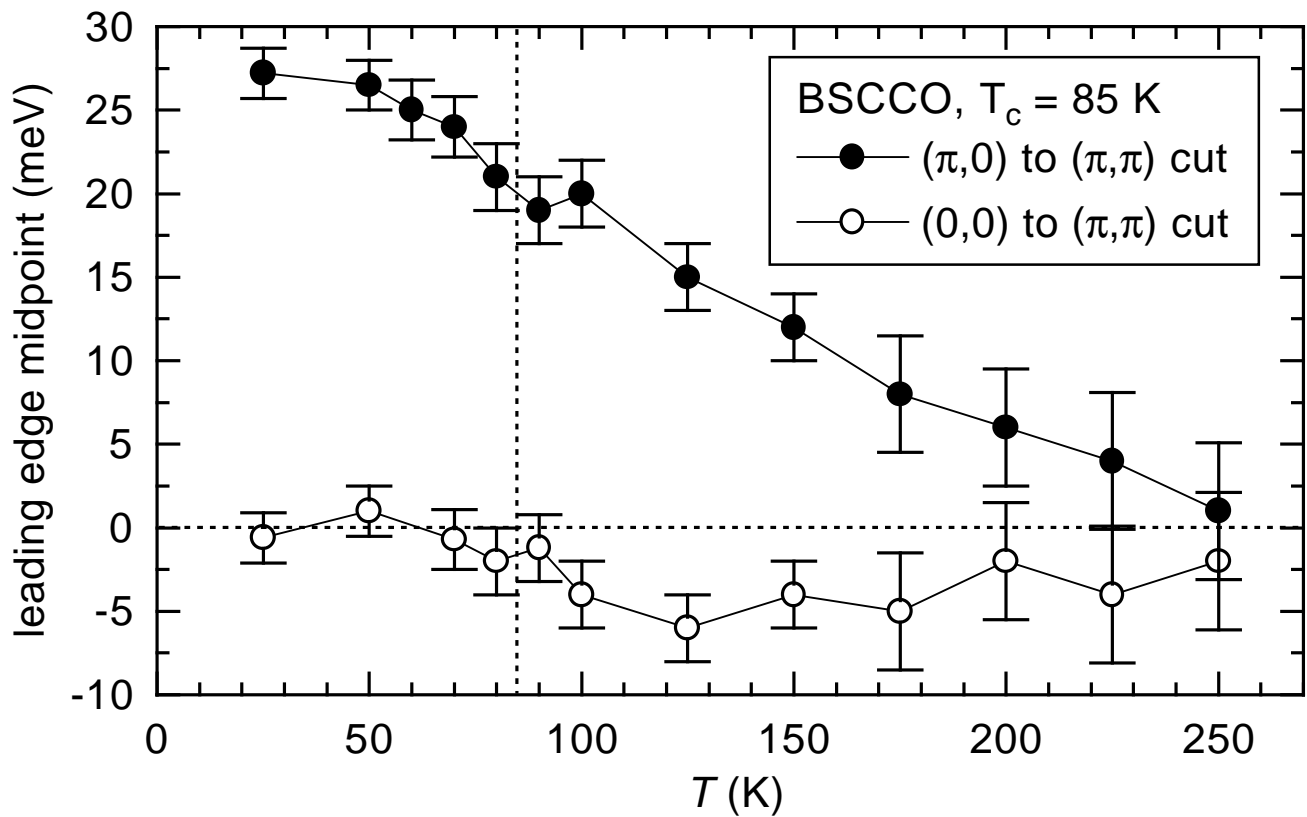
\begin{figure}
\caption{The temperature dependence of leading edge midpoints for
spectra taken at FS crossings near (1,0.2)$\pi$ and 
(0.4,0.4)$\pi$ of an underdoped BSCCO single crystal.  The 
difference between the two represents an energy gap that 
decreases continuously from 28$\pm$2 at 25 K to near zero 
at 225 K. }
\label{fig3}
\end{figure}

The superconducting (13 K) and normal state (75 K) leading 
edge shift vs $k$ curves for the $T_c$ = 
46 K sample agree very well, suggesting the two gaps are 
intimately related.  In the $T_c$ = 78 K 
sample, the anisotropy of the gap is similar above and below
$T_c$, but the magnitude shows a 
marked reduction at 100 and 150 K.  These observations raise two
questions:  Does the gap change continuously through $T_c$, and 
does the gap close at some higher temperature?  To clarify these
issues, more detailed temperature-dependence measurements are 
shown in Fig.\ \ref{fig3}.  The leading edge midpoint energies
on the FS for the maximum (along ($\pi$,0) to
($\pi$,$\pi$)) and minimum gap (along (0,0) to ($\pi$,$\pi$))
are plotted for an underdoped single crystal sample with 
$T_c$ = 85 K.  Measuring the temperature dependence of the 
gap with ARPES requires caution because thermal broadening
and line shape changes may affect the leading edge positions.
To partially cancel these effects, we take the gap to be the
difference between the leading edge midpoints at the two 
$k$-space positions (the leading edge shift).  This gives a
gap of 28$\pm$2 meV ($\sim 20\%$ larger than a thin film 
of the same $T_c$) that changes smoothly
above and below $T_c$, supporting the idea of a single gap 
function evolving with temperature.  The gap becomes 
consistent with zero around 225 K.

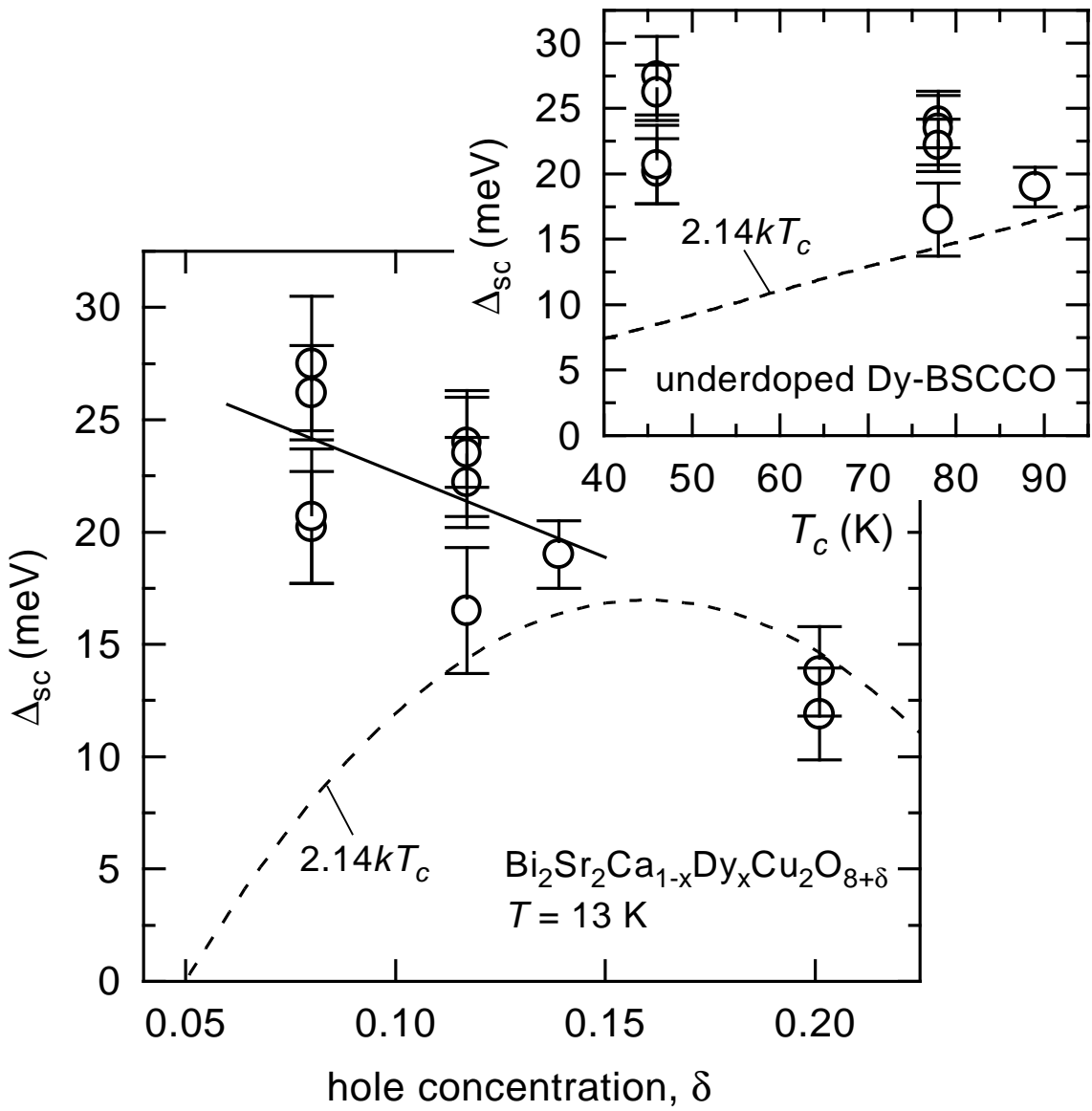
\begin{figure}
\caption{Inset:  The superconducting state gap $\Delta_{sc}$ 
from leading edge shifts measured at 13 K 
on Bi$_2$Sr$_2$Ca$_{1-x}$Dy$_x$Cu$_2$O$_{8+\delta}$ 
plotted vs $T_c$\@.  The nine samples measured came from three 
growth and annealing runs, and therefore fall into three $T_c$ 
groups.  The dashed line is the 
standard BCS mean-field $d$-wave prediction \protect\cite{Won} 
with $\Delta_{sc} = 2.14kT_c$, 
shown to highlight the non-mean-field trend of the data.
Main panel:  $\Delta_{sc}$ vs doping $\delta$, with 
$\delta$ inferred from 
$T_c/T_{c,max}$ \protect\cite{PreslandGroen}.  The energy scale 
from $T_c$ ($2.14kT_c$, dashed line) shows very different 
behavior from the linear fit 
to the $\Delta_{sc}$ data points (straight line) from underdoped 
samples.  By contrast, 
the gap values for overdoped samples \protect\cite{White} decrease 
in the conventional way.
}
\label{fig4}
\end{figure}

Next, we focus on the superconducting state gap $\Delta_{sc}$ as a 
function of $T_c$ 
(Fig.\ \ref{fig4}), using the maximum difference between the 
leading edge midpoints on the 
underlying FS\@.  For these measurements, this 
``internally-referenced'' gap shows the same trend as the 
leading edge midpoint near ($\pi$,0) 
(the {\it d}-wave gap maximum) vs $T_c$\ since the FS crossing
near 0.4($\pi$,$\pi$) shows a gap consistent with zero.  
The nine samples measured fall into three $T_c$ groupings (all 
underdoped) representing three 
thin film crystal growth and annealing runs.  In spite of 
substantial error bars, it is clear that 
the familiar BCS mean-field result of $\Delta_{sc}\propto T_c$ is 
violated 
(for example, $\Delta_{sc}(0) = 2.14kT_c$ for a 
weak-coupling $d_{x^2-y^2}$  solution \cite{Won}).  Instead 
$\Delta_{sc}$ stays constant or
increases slightly as $T_c$ is reduced.

For comparison with phase diagrams, 
Fig.\ \ref{fig4} shows $\Delta_{sc}$  vs doping $\delta$ (hole 
concentration per planar Cu)
for the same underdoped samples, 
with each $T_c$ converted to $\delta$ using the 
empirical relation $T_c/T_{c,max} = 1 - 82.6(\delta - 0.16)^2$ 
\cite{PreslandGroen}.  Once again, the predicted BCS 
weak-coupling {\it d}-wave result, 
$\Delta_{sc} = 2.14kT_c$ (Fig.\ \ref{fig4} dashed line), shows no 
resemblance 
to the observed doping dependence.  By contrast as $T_c$ decreases 
on the overdoped side,
the gap falls rapidly\cite{White}  (although it may not be in
quantitative agreement with the mean-field calculation because
of the {\it ad hoc} leading edge midpoint criterion for the gap).  
The failure of the mean-field theory prediction for $\Delta_{sc}$ 
as a function of $T_c$ (or doping) in the underdoped regime
is our main result; 
it indicates that the $\Delta_{sc}$ and $T_c$ represent two 
distinct energy scales for underdoped BSCCO.
	
The non-mean-field $\Delta_{sc}$ vs $T_c$ behavior is naturally 
related to the similarity of $\Delta$ 
above and below $T_c$  in models with phase fluctuations in the 
complex superconducting 
state order parameter \cite{Emery,Doniach}.  Paired electrons, and 
therefore a gap, exist below a mean-field temperature $T_{MF}$ 
that is proportional to $\Delta$.  The 
zero-resistance $T_c$  is the temperature 
at which long-range phase coherence is established.  $T_{MF}$ and 
$T_c$ are the same in BCS theory
since fluctuations are not considered, but $T_c$ may be lower than 
$T_{MF}$ in underdoped 
cuprates where the phase stiffness is expected to be small.  
Photoemission is insensitive to 
phase, so it measures a gap below $T_{MF}$.  The phase stiffness 
is proportional to the superfluid
density $n_s$.  Thus if the phase stiffness energy scale instead of 
the gap determines $T_c$ , 
then $T_c$ should decrease as the doping is lowered, quite apart 
from any changes in $\Delta$ and 
$T_{MF}$, in agreement with the data of Fig.\ \ref{fig4}.  In this 
model $\Delta$ is determined by the same 
pairing interaction in the superconducting and normal states.  Muon 
spin relaxation ($\mu$SR) 
measurements support the idea that $T_c$ is determined by $n_s$.  
A ``universal curve'' 
with $T_c\propto n_s$ from $\mu$SR has been reported for a large 
number of underdoped cuprate superconductors 
\cite{Uemura}.  Also, magnetoresistance measurements of 60-K 
Y-Ba-Cu-O (underdoped) show a 
Lorentz-force independent contribution persisting to 200 K 
\cite{Harris} that may be consistent with 
phase fluctuations in the order parameter.
	
Various extensions to AndersonÕs original resonating valence bond 
(RVB) idea \cite{PWA} 
give a possible microscopic justification in terms of spin-charge 
separation for the phase 
fluctuation model, including the two energy scales ($kT_c$ and 
$\Delta$) and a distinct pseudogap 
regime \cite{Kotliar,Baskaran,Tanamoto,Wen,Altshuler}.  
Spin-charge separation provides a means of producing
pairing of excitations without superconductivity.  The 
fermionic spin excitations (spinons) pair 
into singlets at a temperature $T_s > T_c$ for underdoped cuprates, 
where $T_s$ is proportional to 
the gap.  $T_s$ is similar to $T_{MF}$ and may represent a 
crossover instead of a true phase 
transition.  Also, $T_s$ decreases with increasing $\delta$ in 
agreement with the trend of the upper line in 
Fig.\ \ref{fig4}.  The pairing was predicted to be {\it d}-wave, 
another area of agreement with the data.  
Because of the apparent flattening of the normal state gap 
near the {\it d}-wave node position 
(Fig.\ \ref{fig2}), the data could also be consistent with 
recent work predicting the pseudogap 
regime to be a mixture of {\it d}-wave spinon pairing and pockets 
of spinon FS\@\cite{Wen}.  
In either case, the charge excitations (holons) Bose condense at 
$T_c$, so once again $T_c\propto n_s$, 
the two-dimensional Bose condensation result.  
$T_c$ is also the temperature at which phase coherence 
between the singlet pairs appears.  These models are in agreement 
with the ARPES data in the 
doping dependence of the gap, and in the existence of {\it d}-wave 
pairing well above $T_c$ for 
underdoped samples.
	
In summary, the increase in the superconducting state gap with 
decreasing $T_c$ violates 
the BCS mean-field theory prediction and suggests the existence of 
an energy scale for 
pairing that is separate from, and higher than, $kT_c$.  This 
energy scale accounts for the 
pseudogap above $T_c$ (normal state gap) seen by a variety of 
experimental probes in 
underdoped cuprate superconductors.  As measured by ARPES, the 
normal state gap is highly 
anisotropic, and it is similar in magnitude and 
{\it k}-dependence to the superconducting state 
gap, supporting the idea of a common underlying pairing 
interaction. 

\acknowledgments

We acknowledge very helpful discussions with S. A. Kivelson,
P. W. Anderson and S. Doniach.
SSRL is operated by the DOE Office of 
Basic Energy Sciences, Division 
of Chemical Sciences.  This work was supported by ONR Grant
No.\ N00014-95-1-0760; the 
Division of Materials 
Science, DOE; and NSF Grant No.\ DMR 9311566.  
The research at Varian was supported in part by NRL and ONR.

\begin{references}
\bibitem{filmnote}In particular, thin films and bulk single 
crystals give very similar gap anisotropy.  However, thin
films may show less strongly peaked normal state line shapes
and slightly smaller gap values than bulk single crystals.
\end{references}
\end{document}